\documentclass[aps,pre,twocolumn,nofootinbib,floatfix]{revtex4}
\usepackage{amsmath}
\usepackage{epsfig}

\input{tcilatex}

\begin{document}

\title{Decay rate and renormalized frequency shift of superradiant exciton
in a cylindrical quantum wire}
\author{Yueh-Nan Chen}
\email{ynchen.ep87g@nctu.edu.tw}
\author{Der-San Chuu}
\email{dschuu@cc.nctu.edu.tw}
\affiliation{Department of Electrophysics, National Chiao-Tung University, Hsinchu 30050,
Taiwan}
\date{\today }

\begin{abstract}
The decay rate and renormalized frequency shift of superradiant exciton in a
cylindrical quantum wire are studied. The transition behavior from 1D wire
to 2D film is examined through the property of the radiative decay. Similar
to the case in a quantum well, the decay rate of the higher mode exciton is
larger than that of the lower mode one. Moreover, it is also found the decay
rate and frequency shift do not show oscillatory dependence on wire radius
because of the conservation of angular momentum.

PACS: 42.50.Fx, 32.70.Jz, 71.35.-y, 71.45.-d
\end{abstract}

\maketitle

\address{Department of Electrophysics, National Chiao Tung University,
Hsinchu 30050, Taiwan}

\address{Department of Electrophysics, National Chiao Tung University,
Hsinchu 30050, Taiwan}

\address{Department of Electrophysics, National Chiao Tung University,
Hsinchu 30050, Taiwan}

\address{Department of Electrophysics, National Chiao Tung University,
Hsinchu 300, Taiwan}





Since Dicke\cite{1} pointed out the concept of superradiance, the coherent
effect for spontaneous radiation of various systems has attracted extensive
interest both theoretically and experimentally \cite{2,3,4,5}. In bulk
crystal,\textit{\ }the excitons will couple with photons to form polaritons%
\cite{6}-mixed modes in which energy oscillates back and forth between the
exciton and the radiation field. What makes the excitons trapped in the bulk
crystal is the conservation of crystal momentum. If one considers a thin film%
\cite{7}, the excitons can undergo radiative decay as a result of the broken
crystal symmetry along the normal direction of the film plane. The decay
rate of excitons in a thin film is enhanced by a factor of $(\lambda /d)^{2}$
compared to a lone exciton in an empty lattice, where $\lambda $ is the wave
length of emitted photon and $d$ is the lattice constant of the film.

Lots of investigations on the radiative linewidth of excitons in quantum
wells have been performed. An abnormal increase of excitonic radiative
lifetime with the decrease of well width below 5$nm$ for \hspace{0.06in}In$%
_{x}$Ga$_{1-x}$As/InP quantum well was observed by Cebulla \emph{et al}.\cite%
{8}. Brandt \emph{et al}.\cite{9} measured the radiative lifetime of
excitons in InAs quantum sheets and observed the increasing of radiative
lifetime with the decreasing of well thickness. Hanamura\cite{10}
investigated theoretically the radiative decay rate of quantum dot and
quantum well. The obtained results are in agreement with that of Lee and
Liu's\cite{7} prediction for thin films. Knoester\cite{11} studied the
radiative dynamics crossover from the small thickness, superradiant exciton
regime to bulk crystal, polariton regime. The oscillating dependence of the
radiative width of the exciton-like polaritons with the lowest energy on the
crystal thickness was found. Recently, G. Bj\"{o}rk \emph{et al}.\cite{12}
examined the relationship between atomic and excitonic superradiance in thin
and thick slab geometries. They demonstrated that superradiance can be
treated by a unified formalism for atoms, Frenkel excitons, and Wannier
excitons. In V. M. Agranovich \emph{et al}.'s work\cite{13}, a detailed
microscopic study of Frenkel exciton-polariton in crystal slabs of arbitrary
thickness was performed.

For lower dimensional systems, the decay rate of the exciton is enhanced by
a factor of $\lambda /d$ in a linear chain\cite{14}. First observation of
superradiant short lifetimes of excitons was performed by Ya. Aaviksoo \emph{%
et al}.\cite{15} on surface states of the anthracene crystal. A. L. Ivanov
and H. Haug\cite{16} predicted the existence of exciton crystal, which
favors coherent emission in the form of superradiance, in quantum wires. Y.
Manabe \emph{et al}.\cite{17} considered the superradiance of interacting
Frenkel excitons in a linear chain. Recently, we have also shown the
superradiant decay of the quantum wire exciton is greatly enhanced in a
planar microcavity\cite{18}. In this paper, the exciton is assumed to be
confined in a hollow cylindrical quantum wire. The crossover of the
superradiant decay rate of the exciton from a wire with small radius to the
2D limit is explicitly obtained. Moreover, the crossover of the coherent
frequency shift from narrow wire to thin film is also investigated by using
the method of renormalization\cite{19,20}.

We consider a free exciton in a hollow cylindrical quantum wire lying on the 
$z$ axis with simple cubic structure (Fig. 1). For the physical phenomenon
we are interested in, we shall concentrate on the investigations of
semiconductor quantum wires rather than existing carbon wires \cite{21,22}
since the wavelength of the emitted photon is much larger than the diameter
of the single-wall carbon tubes. The crossover behavior from 1D wire to 2D
film may not be examined on carbon systems. However, we hope our model can
serve as a first step toward the understanding of the exciton decay in
carbon nanotubes. 
\begin{figure}[h]
\includegraphics[width=8cm]{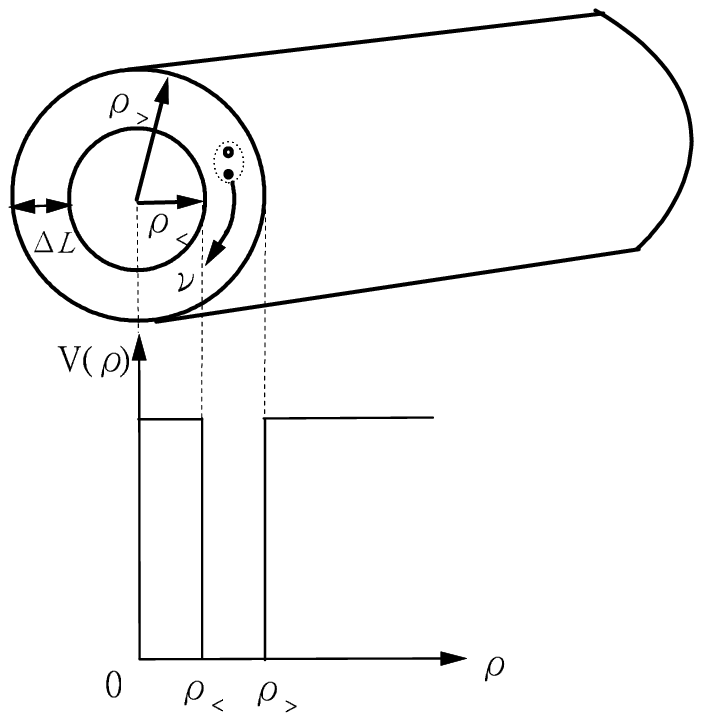}
\caption{Schematic view of the quantum wire structure and its defining
potential profile.}
\end{figure}

If the difference between the inner radius($\rho _{<}$) and outer radius($%
\rho _{>}$) is much smaller than the effective Bohr radius of the exciton,
i. e. $\Delta L<<a_{ex}$, one may approximate that the exciton is confined
on the cylindrical surface with radius $\rho $($\approx \rho _{<}\approx
\rho _{>}$). This means the exciton is trapped in an infinite deep and
narrow quantum wire. The radius $\rho $ is about $Nd/2\pi $, where $N$ is
the number of lattice points in the circumference direction and $d$ is the
lattice spacing. If one further assumes the radius of the wire is much
larger than the effective Bohr radius of the exciton, variations of the wire
radius only cause few changes on the Wannier exciton wavefunction. In this
case, the main contributions to the \emph{superradiant} decay rate and
frequency shift still come from the bandgap energy and the number of the
lattice points within a wavelength of the emitted photon\cite{20}.
Therefore, one can first consider the exciton as a particle with angular
momentum $\nu $ and longitudinal momentum $k_{z}$. After figuring out the
decay rate and frequency shift of a Frenkel exciton, the corresponding ones
of a Wannier Exciton can be obtained by replacing the single-atom dipole
matrix element $\mathbf{\chi }$ with the effective dipole matrix element\cite%
{7}. Thus, the Hamiltonian for the exciton is

\begin{equation}
H_{ex}=\sum_{\nu k_{z}}E_{\nu k_{z}}c_{\nu k_{z}}^{\dagger }c_{\nu k_{z}},
\end{equation}
where $c_{\nu k_{z}}^{\dagger }$ and $c_{\nu k_{z}}$ are the creation and
destruction operators of the exciton, respectively. The Hamiltonian of free
photon is

\begin{equation}
H_{ph}=\sum_{q^{\prime }\nu ^{\prime }k_{z}^{\prime }}\hbar c(q^{\prime
2}+k_{z}^{\prime 2})^{1/2}b_{q^{\prime }\nu ^{\prime }k_{z}^{\prime
}}^{\dagger }b_{q^{\prime }\nu ^{\prime }k_{z}^{\prime }},
\end{equation}%
where $b_{q^{\prime }\nu ^{\prime }k_{z}^{\prime }}^{\dagger }$ and $%
b_{q^{\prime }\nu ^{\prime }k_{z}^{\prime }}$ are, respectively, the
creation and destruction operators of the photon. The wave vector $\mathbf{k}%
^{\prime }$\hspace{0.06in}of the photon is separated into two parts: $%
k_{z}^{\prime }$ is the parallel component of $\mathbf{k}^{\prime }$ on the $%
z$ direction such that $k^{\prime 2}=q^{\prime 2}+k_{z}^{\prime 2}$.

The interaction between the exciton and the photon can be expressed as

\begin{eqnarray}
H^{\prime } &=&\sum_{i}\sum_{q^{\prime }\nu ^{\prime }k_{z}^{\prime }}\frac{e%
}{mc}\sqrt{\frac{2\pi \hbar c}{(q^{\prime 2}+k_{z}^{\prime 2})^{1/2}v}}(%
\mathbf{\epsilon }_{q^{\prime }\nu ^{\prime }k_{z}^{\prime }}\cdot \mathbf{p}%
_{i})  \notag \\
&&\times \lbrack b_{q^{\prime }\nu ^{\prime }k_{z}^{\prime }}^{\dagger
}H_{\nu ^{\prime }}^{(1)}(q^{\prime }\rho )\exp (i\nu ^{\prime }\varphi
_{i}+ik_{z}^{\prime }l_{i})+\mathbf{h.c}.],
\end{eqnarray}%
where $m$ is the electron mass, ($\rho _{i},\varphi _{i},l_{i}$) is the
position of the electron $i$ in the cylindrical wire, $\mathbf{p}_{i}$ is
the corresponding momentum operator of the electron $i$\ , $\mathbf{\epsilon 
}_{q^{\prime }\nu ^{\prime }k_{z}^{\prime }}$ is the polarization vector of
the photon, and $H_{\nu ^{\prime }}^{(1)}(q^{\prime }\rho )$ is the Hankel
function.

The essential quantity involved is the matrix element of $H^{\prime }$
between the ground state $\left| G\right\rangle $ and the exciton state $%
\left| \nu ,k_{z}\right\rangle $. We know that the interaction matrix
elements of $H^{\prime }$ can be written as

\begin{equation}
\left\langle \nu ,k_{z}\right| H^{\prime }\left| G\right\rangle
=\sum_{l,\varphi }\left\langle c,(l,\varphi );v,(l,\varphi )\right| U_{\nu
k_{z}}^{\ast }(l,\varphi )H^{\prime }\left| G\right\rangle ,
\end{equation}
because the exciton state can be expressed as

\begin{equation}
\left| \nu ,k_{z}\right\rangle =\sum_{l,\varphi }U_{\nu k_{z}}^{\ast
}(l,\varphi )\left| c,(l,\varphi );v,(l,\varphi )\right\rangle ,
\end{equation}
in which the excited state $\left| c,(l,\varphi );v,(l,\varphi
)\right\rangle $ is defined as

\begin{equation}
\left| c,(l,\varphi );v,(l,\varphi )\right\rangle =a_{c,(l,\varphi
)}^{\dagger }a_{v,(l,\varphi )}\left| G\right\rangle ,
\end{equation}
where $a_{c,(l,\varphi )}^{\dagger }(a_{v,(l,\varphi )})$ is the creation
(destruction) operator of an electron in the conduction (valence) band at
lattice site $(l,\varphi ).$ The expansion coefficient $U_{\nu k_{z}}^{\ast
}(l,\varphi )$ is the exciton wave function in the cylindrical tubule:

\begin{equation}
U_{\nu k_{z}}^{\ast }(l,\varphi )=\frac{1}{\sqrt{N}}\frac{1}{\sqrt{N^{\prime
}}}e^{i\nu \varphi +ik_{z}l},
\end{equation}
where the coefficient $1/\sqrt{N^{\prime }}$ is for the normalization of the
state $\left| \nu ,k_{z}\right\rangle $.

After summing over $l$ and $\varphi $ in Eq. (4), we have

\begin{eqnarray}
\left\langle \nu ,k_{z}\right| H^{\prime }\left| G\right\rangle
&=&\sum_{q^{\prime }gn_{\nu }}\frac{e}{mc}\sqrt{\frac{2\pi \hbar c}{%
(q^{\prime 2}+(k_{z}+g)^{2})^{1/2}v}}\times  \notag \\
&&[b_{q^{\prime }\mathbf{,\nu +}n_{\nu }\mathbf{,}k_{z}+g}(\mathbf{\epsilon }%
_{q^{\prime },\nu +n_{\nu },k_{z}+g}\cdot  \notag \\
&&\mathbf{A}_{\nu +n_{\nu },k_{z}+g})H_{\nu +n_{\nu }}^{(1)}(q^{\prime }\rho
)+\mathbf{h.c}.],
\end{eqnarray}%
where

\begin{eqnarray}
\mathbf{A}_{\nu +n_{\nu },k_{z}+g} &=&\sqrt{NN^{\prime }}\int d^{2}\mathbf{%
\tau }\exp (i(k_{z}+g)\tau _{z}+i(\nu +n_{\nu })\tau _{\varphi })  \notag \\
&&\times w_{c}(\mathbf{\tau }_{z}\mathbf{,}\tau _{\varphi })(-i\hbar \mathbf{%
\nabla })w_{v}(\mathbf{\tau }_{z}\mathbf{,}\tau _{\varphi }),
\end{eqnarray}%
, $w_{c}(\mathbf{\tau }_{z}\mathbf{,}\tau _{\varphi })$ and $w_{v}(\mathbf{%
\tau }_{z}\mathbf{,}\tau _{\varphi })$ are, respectively, the Wannier
functions for the conduction band and the valence band at site 0, and $g$ $%
(n_{\nu })$ is the reciprocal lattice in $k_{z}$ ($k_{\varphi }$) direction.
Hence the interaction between the exciton and the photon (in the resonance
approximation) can be written in the form

\begin{equation}
H^{\prime }=\sum_{n_{\nu }g}\sum_{q^{\prime }\nu k_{z}}D_{q^{\prime },\nu
+n_{\nu },k_{z}+g}b_{q^{\prime },\nu +n_{\nu },k_{z}+g}c_{\nu
k_{z}}^{\dagger }+\mathbf{h.c.,}
\end{equation}%
where

\begin{eqnarray}
D_{q^{\prime },\nu +n_{\nu },k_{z}+g} &=&H_{\nu +n_{\nu }}^{(1)}(q^{\prime
}\rho )\mathbf{\epsilon }_{q^{\prime },\nu +n_{\nu },k_{z}+g}\cdot \mathbf{A}%
_{\nu +n_{\nu },k_{z}+g}  \notag \\
&&\times \frac{e}{mc}\sqrt{\frac{2\pi \hbar c}{(q^{\prime
2}+(k_{z}+g)^{2})^{1/2}v}}.
\end{eqnarray}

Now, we assume that at time $t=0$ the exciton is in the mode $\nu ,k_{z}.$
For time $t>0$,\hspace{0.06in}the state $\left| \psi (t)\right\rangle $%
\hspace{0.06in}can be written as

\begin{eqnarray}
\left| \psi (t)\right\rangle &=&f_{0}(t)\left| \nu ,k_{z};0\right\rangle \\
&&+\sum_{q^{\prime }n_{\nu }g}f_{G;q^{\prime },\nu +n_{\nu
},k_{z}+g}(t)\left| G;q^{\prime },\nu +n_{\nu },k_{z}+g\right\rangle , 
\notag
\end{eqnarray}%
where $\left| \nu ,k_{z};0\right\rangle $ is the state with an exciton in
the mode $\nu ,k_{z}$ in the cylindrical quantum wire, $\left| G;q^{\prime
},\nu +n_{\nu },k_{z}+g\right\rangle $ represents the state in which the
electron-hole pair recombines and a photon in the mode $q^{\prime },\nu
+n_{\nu },k_{z}+g$ is created, and $f_{0}(t)$ and $f_{G;q^{\prime },\nu
+n_{\nu },k_{z}+g}(t)$ are, respectively, the probability amplitudes of the
state $\left| \nu ,k_{z};0\right\rangle $ and $\left| G;q^{\prime },\nu
+n_{\nu },k_{z}+g\right\rangle $.

In the resonance approximation, the probability amplitude $f_{0}(t)$ can be
expressed as\cite{7}

\begin{equation}
f_{0}(t)=\exp (-i\Omega _{\nu k_{z}}t-\frac{1}{2}\gamma _{\nu k_{z}}t),
\end{equation}%
where

\begin{equation}
\gamma _{\nu k_{z}}=2\pi \sum_{q^{\prime }n_{\nu }g}\left| D_{q^{\prime
},\nu +n_{\nu },k_{z}+g}\right| ^{2}\delta (\omega _{q^{\prime },\nu +n_{\nu
},k_{z}+g})
\end{equation}%
and

\begin{equation}
\Omega _{\nu k_{z}}=\mathcal{P}\sum_{q^{\prime }n_{\nu }g}\frac{\left|
D_{q^{\prime },\nu +n_{\nu },k_{z}+g}\right| ^{2}}{\omega _{q^{\prime },\nu
+n_{\nu },k_{z}+g}}
\end{equation}%
with $\omega _{q^{\prime },\nu +n_{\nu },k_{z}+g}=E_{\nu +n_{\nu
},k_{z}+g}/\hbar -c\sqrt{q^{\prime 2}+(k_{z}+g)^{2}}.$ Here $\gamma _{_{\nu
k_{z}}}$ and $\Omega _{\nu k_{z}}$ are, respectively, the decay rate and
frequency shift of the exciton. And $\mathcal{P}$ means the principal value
of the integral.

If we neglect the Umklapp process, the exciton decay rate in the optical
region can be calculated straightforwardly and is given by

\begin{equation}
\gamma _{\nu k_{z}}=\QATOPD\{ . {\frac{3}{2}\pi ^{2}\gamma _{0}\frac{\rho }{d%
}\frac{1}{k_{0}d}\left| H_{\nu }^{(1)}(\rho \sqrt{k_{\nu }^{2}-k_{z}^{2}}%
)\right| ^{2},\text{ \hspace{0.06in}}k_{z}<k_{\nu }}{0,\text{ \ \ \ \ \ \ \
\ \ \ \ \ \ \ \ \ otherwise \ \ \ \ \ \ \ \ \ }},
\end{equation}%
where $k_{\nu }=E_{k_{z}\nu }/\hbar =k_{0}+\frac{\hbar ^{2}\nu ^{2}}{2\mu
\rho ^{2}}$,

\begin{equation}
\gamma _{0}=\frac{4e^{2}\hbar k_{0}}{3m^{2}c^{2}}\left| \mathbf{\chi }%
\right| ^{2},
\end{equation}%
and

\begin{equation}
\mathbf{\chi }=\int d^{2}\mathbf{\tau }w_{c}(\mathbf{\tau }_{z}\mathbf{,}%
\tau _{\varphi })(-i\hbar \mathbf{\nabla })w_{v}(\mathbf{\tau }_{z}\mathbf{,}%
\tau _{\varphi }).
\end{equation}

Here, $\mu $ is the effective mass of the exciton, $\mathbf{\chi }^{\ast }$
represents the single-atom dipole matrix element for an electron jumping
from the excited state in the conduction band back to the hole state in the
valence band, and $\gamma _{0}$ is the decay rate of an isolated atom. We
see from Eq. (16) that $\gamma _{_{k_{z}n}}$ is proportional to $1/(k_{0}d)$
and $\rho /d$. This is just the superradiance factor coming from the
coherent contributions of atoms. If $k_{z}$ is larger than $k_{\nu }$, these
excitonic modes (trapped modes) are not capable of radiative decay. This is
simply because the energy $\hbar ck_{\nu }$ of that exciton is not
sufficient to produce a photon.

\begin{figure}[th]
\includegraphics[width=8cm]{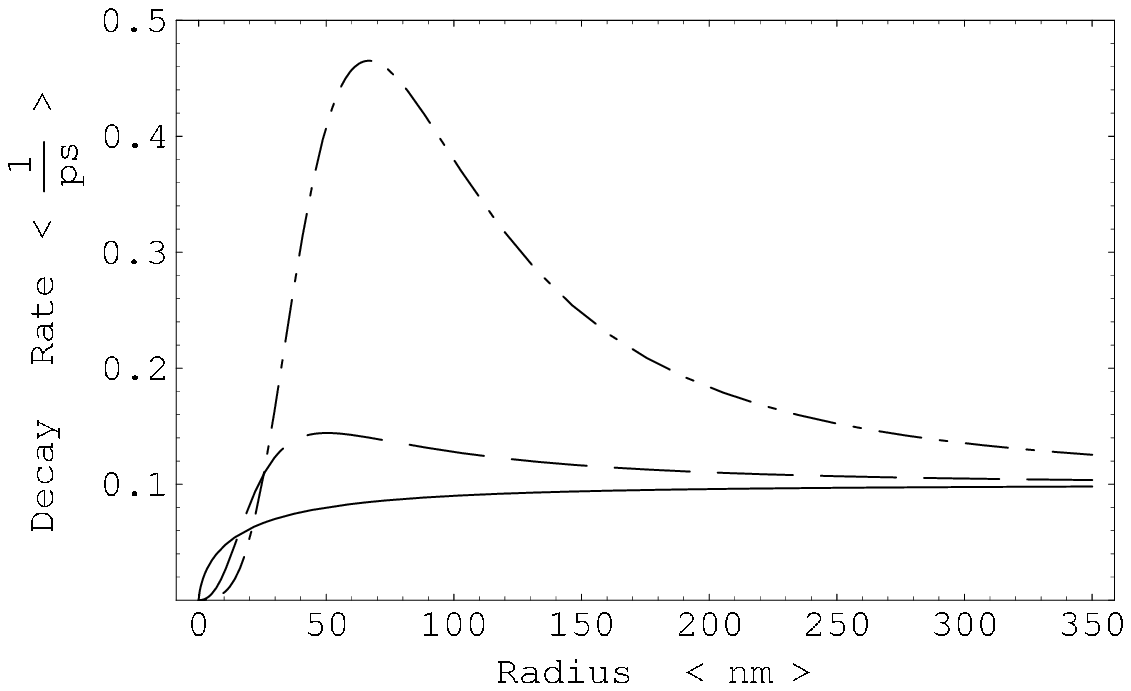}
\caption{Decay Rate of the superradiant exciton as a function of the wire
radius. The vertical and horizontal units are ps$^{-1}$ and $nm,$
respectively. And the solid, dashed, and dot-dashed lines represnet the $%
\protect\nu =0,1,$ and $2$ modes, respectively. In this and following graph
the paramaters of a GaAs quantum well are chosen to obtain the numerical
results.}
\end{figure}

In Fig. 2 we plot the decay rate $\gamma _{\nu k_{z}}$ as a function of wire
radius. Our numerical results are obtained by employing the data of GaAs
(band gap is 1.52 eV). For $\nu =0$ mode(solid line), the decay rate
increases with the increasing of wire radius and approaches 2D limit as $%
\rho $ is large comparing to the wavelength of the emitted photon in a 2D
thin film. For higher modes($\nu =1,2$), the results depend on the kinetic
energy $\frac{\hbar ^{2}\nu ^{2}}{2\mu \rho ^{2}}$. As can be seen from the
figure, the decay rate also approaches 2D limit for large radius. On the
other hand, the decay rate increases as the wire radius is decreased until
the maximum decay rate is attained. After that a further reducing of the
radius leads to a sharp decrease of the decay rate. This is similar to the
transition from 2D to 3D: the higher wave number modes have larger maximum
decay rate\cite{12}. One also notes in small radius regime, the decay rate
of the higher mode is smaller than that of the lower mode, i.e. $\gamma
_{\nu =2,k_{z}}>$ $\gamma _{\nu =1,k_{z}}>\gamma _{\nu =0,k_{z}}$. This
phenomenon is also found in the quantum well system, and can be ascribed to
the effects of interference\cite{20}. Another interesting result is that the
decay rate does not show any oscillatory dependence on radius while it shows
oscillatory dependence on layer thickness in the case of the semiconductor
thin film\cite{11,20}. The reason is that the angular momentum of the
exciton is conserved in a cylindrical system, but, in a semiconductor thin
film, the momentum is not conserved in the direction of broken symmetry. If
we sum over all the sites in Eq. (4), the phase will run through a circle
and thus preserve the same value.\ 

One might suspect the practical value of our numerical results because our
model is based on the one-layer assumption and the idealized assumptions ,
e.g. perfect cylindrical symmetry and infinite confinement. However, as we
mentioned above, for the superradiant decay rate and frequency shift one
only needs to replace the dipole matrix element with an effective one. This
is why our result agrees well with that of a realistic GaAs quantum well in
the large radius limit\cite{10}. On the other hand, such an assumption may
slightly deviate from the real case as the radius is small (e.g. as the
radius approaches $a_{ex})$. This is because the wavefunction of the exciton
becomes more compact in the quantum wire with small radius limit, and it
causes the increasing of the dipole matrix element $\mathbf{\chi }$ in Eq.
(17). Besides, the quality of the quantum wire also influences the
observation of the crossover, i.e. the thickness fluctuations should be
controlled well, otherwise it will destroy the coherence in the circular
direction.

Now let us turn to the results for the renormalized frequency shift. The
frequency shift in Eq. (15) can be expressed as

\begin{eqnarray}
\Omega _{\nu k_{z}} &=&\frac{\pi e^{2}\hbar }{m^{2}c^{2}v}\mathcal{P}%
\sum_{q^{\prime }n_{\nu }g}\frac{\left| \mathbf{\epsilon }_{q^{\prime },\nu
+n_{\nu },k_{z}+g}\cdot \mathbf{A}_{\nu +n_{\nu },k_{z}+g}\right| ^{2}}{%
\sqrt{q^{\prime 2}+(k_{z}+g)^{2}}}  \notag \\
&&\times \frac{\left| H_{\nu +n_{\nu }}^{(1)}(q^{\prime }\rho )\right| ^{2}}{%
E_{\nu +n_{\nu },k_{z}+g}/(\hbar c)-\sqrt{q^{\prime 2}+(k_{z}+g)^{2}}}.
\end{eqnarray}

As seen from above, the frequency shift suffers from ultraviolet divergence
when $g$ and $n_{\nu }$ are large, and has infrared divergence when the
denominator approaches zero. Following the procedure as shown in the work of
Lee, Chuu, and Mei\cite{19}, the divergent problem is solved by
renormalization.

In the usual renormalization procedure used in quantum field theory, the
infinite quantities in Green function is subtracted by some infinite
quantities to make them finite. The subtraction procedure is possible by
substituting some finite physical quantities, such as renormalized masses,
charges, and wave functions. The finite physical quantities observed from
bare infinite quantities can be visualized, in the case of bare charge for
example, as the polarization of vacuum. This is equivalent to say that there
is virtual photon and opposite charged electron cloud surrounding the bare
charge making the charge measured from outside become finite--a phenomena
similar to that of shielding in dielectric material. But the renormalization
procedure in quantum field theory is for free electrons. In the case of the
condensed matter, the electrons are confined by periodic potential and
complex interactions. Borrowing from the concepts of renormalization used in
quantum field theory, we have

\begin{equation}
\Omega _{\nu k_{z}}^{ren}=\Omega _{\nu k_{z}}-\underset{k_{0}\rightarrow
0,d\rightarrow \infty }{\lim }\Omega _{\nu k_{z}},
\end{equation}%
where the two limiting processes $k_{0}\rightarrow 0$ and $d\rightarrow
\infty $ reduce the exciton to a free electron. In the limiting process $%
d\rightarrow \infty $, the ordinary exciton becomes a lone exciton standing
alone in an empty lattice with no interaction with other atoms. And the
process $k_{0}\rightarrow 0$\ means that\ there is no energy difference
between electron and hole.

We will now show that the ultraviolet divergence comes from the inclusion of
Umklapp process. Define

\begin{equation}
\Omega _{\nu k_{z}}(l,\varphi )=\sum_{k_{2}n_{\varphi }}J_{\nu
k_{z}}(k_{2},n_{\varphi })\exp (ik_{2}l+in_{\varphi }\varphi )
\end{equation}%
with

\begin{eqnarray}
J_{\nu k_{z}}(k_{2},n_{\varphi }) &=&\frac{\pi e^{2}\hbar }{m^{2}c^{2}v}%
\mathcal{P}\sum_{q^{\prime }}\frac{\left| \mathbf{\epsilon }_{q^{\prime
}n_{\varphi }k_{2}}\cdot \mathbf{\chi }\right| ^{2}}{\sqrt{q^{\prime
2}+k_{2}^{2}}}  \notag \\
&&\times \frac{\left| H_{n_{\varphi }}^{(1)}(q^{\prime }\rho )\right| ^{2}}{%
E_{n_{\varphi }k_{2}}/(\hbar c)-\sqrt{q^{\prime 2}+k_{2}^{2}}}.
\end{eqnarray}

From the above equations, we have

\begin{eqnarray}
&&\sum_{l,\varphi }\Omega _{\nu k_{z}}(l,\varphi )\exp (-ik_{z}l-i\nu
\varphi )  \notag \\
&=&\sum_{k_{2}n_{\varphi }}J_{\nu k_{z}}(k_{2},n_{\varphi })\sum_{l\varphi
}e^{i(k_{2}-k_{z})l+i(n_{\varphi }-\nu )\varphi }  \notag \\
&=&N^{\prime }N\sum_{gn_{\nu }}J_{\nu k_{z}}(k_{z}+g,\nu +n_{\nu })  \notag
\\
&=&\Omega _{\nu k_{z}}.
\end{eqnarray}

And from Eq. (21) we have lim$_{d\rightarrow 0}\Omega _{\nu k_{z}}=\Omega
_{\nu k_{z}}(l=0,\varphi =0)$, so we can rewrite the renormalized frequency
shift $\Omega _{\nu k_{z}}^{ren}$ as

\begin{equation}
\Omega _{\nu k_{z}}^{ren}=\Omega _{\nu k_{z}}^{ren}(0,0)+\Omega _{\nu
k_{z}}^{coh}
\end{equation}%
with

\begin{equation}
\Omega _{\nu k_{z}}^{ren}(0,0)=\Omega _{\nu
k_{z}}(0,0)-\lim_{k_{0}\rightarrow 0}\Omega _{\nu k_{z}}(0,0)
\end{equation}%
and

\begin{equation}
\Omega _{\nu k_{z}}^{coh}=\sum_{\substack{ l\neq 0  \\ \varphi \neq 0 }}%
e^{-i(k_{z}l+\nu \varphi )}\Omega _{\nu k_{z}}(l,\varphi ).
\end{equation}%
\ \ \ 

As can be seen from Eq. (25), the renormalization affects only the part $%
\Omega _{\nu k_{z}}(0,0)$--the frequency shift of the lone exciton in an
empty lattice--and thus nothing to do with the correlation within the
crystals. The coherent part $\Omega _{\nu k_{z}}^{coh}$ in Eqs. (24) and
(26) is not touched by renormalization procedure. The separation of $\Omega
_{\nu k_{z}}$ into two parts as shown in Eq. (24) is conceptually equivalent
to singling out of the source term of quantum electrodynamical divergence in
a correlated system.

Now we will investigate the origin of ultraviolet divergence. From Eq. (19)
with the substitution $\sum_{q^{\prime }}\rightarrow \int \frac{q^{\prime
}dq^{\prime }}{(2\pi /R^{2})}$ and $\sum_{g}\rightarrow \int \frac{dg}{(2\pi
/L_{z})},$ \ we have

\begin{eqnarray}
\Omega _{\nu k_{z}} &\simeq &\frac{e^{2}\hbar N^{\prime }}{4\pi m^{2}c^{2}}%
\mathcal{P}\sum_{n_{\nu }}\int q^{\prime }dq^{\prime }\int dg\frac{\left| 
\mathbf{\epsilon }_{q^{\prime },\nu +n_{\nu },k_{z}+g}\cdot \mathbf{\chi }%
\right| ^{2}}{\sqrt{q^{\prime 2}+(k_{z}+g)^{2}}}  \notag \\
&&\times \frac{\left| H_{\nu +n_{\nu }}^{(1)}(q^{\prime }\rho )\right| ^{2}}{%
E_{\nu +n_{\nu },k_{z}+g}/(\hbar c)-\sqrt{q^{\prime 2}+(k_{z}+g)^{2}}}.
\end{eqnarray}%
It is noted that the result in Eq. (27) is the same as $\Omega _{\nu
k_{z}}(0,0)$ from Eqs. (21) and (23). Hence we conclude that the ultraviolet
divergence of $\Omega _{\nu k_{z}}(0,0)$ is really the same as that from
Umklapp process of large $g$ and $n_{\nu \text{ }}$that contribute to the
full $\Omega _{\nu k_{z}}.$ Once $\Omega _{\nu k_{z}}(0,0)$ of the lone
exciton is rendered finite by renormalization via Eq. (25), the Umklapp
processes of large $g$ and $n$ that arise from the unrestricted sum over $l$
and $\varphi $ will also be rendered finite simultaneously. Accordingly, $%
\Omega _{\nu k_{z}}^{ren}$ can be written as

\begin{equation}
\Omega _{\nu k_{z}}^{ren}=\frac{\pi e^{2}\hbar N^{\prime }}{m^{2}c^{2}v}%
\mathcal{P}\sum_{q^{\prime }}\frac{\left| \mathbf{\epsilon }_{q^{\prime }\nu
k_{z}}\cdot \mathbf{\chi }\right| ^{2}}{\sqrt{q^{\prime 2}+k_{z}^{2}}}\frac{%
\left| H_{\nu }^{(1)}(q^{\prime }\rho )\right| ^{2}}{E_{\nu k_{z}}/(\hbar c)-%
\sqrt{q^{\prime 2}+k_{z}^{2}}}.
\end{equation}%
and thus the ultraviolet divergence problem is solved.

In the $k_{z}\mathbf{\sim }0$ and $\nu =0$ mode, the renormalized result can
be reduced as

\begin{equation}
\Omega _{\nu k_{z}}^{ren}=\frac{\pi e^{2}\hbar N^{\prime }}{m^{2}c^{2}v}%
\mathcal{P}\sum_{q^{\prime }}\frac{\left| \mathbf{\epsilon }_{q^{\prime }\nu
k_{z}}\cdot \mathbf{\chi }\right| ^{2}}{q^{\prime }}\frac{\left|
H_{0}^{(1)}(q^{\prime }\rho )\right| ^{2}}{k_{0}-q^{\prime }}.
\end{equation}

As seen from above, the frequency shift suffers from infrared divergence
when $q^{\prime }\sim 0$ or $q^{\prime }\sim k_{0}.$ This can be overcome by
substituting $-i\hbar \nabla $ by $-imcq^{\prime }\tau $(Ref. 19) in Eq.
(18) when $q^{\prime }$ is small. It is equivalent to the dipole interaction
form, $H^{\prime }\sim \mathbf{r}\cdot \mathbf{E.}$ With this treatment, we
have

\begin{equation}
\Omega _{\nu k_{z}}^{ren}\sim \frac{2\pi \rho }{d}N^{\prime }\mathcal{P}%
\sum_{q^{\prime }}\mathbf{B}_{q^{\prime }\nu k_{z}}\frac{\left|
H_{0}^{(1)}(q^{\prime }\rho )\right| ^{2}}{k_{0}-q^{\prime }}
\end{equation}%
with

\begin{equation}
\mathbf{B}_{q^{\prime }\nu k_{z}}=\QATOPD\{ . {\frac{\pi e^{2}\hbar }{%
m^{2}c^{2}v}\left| \mathbf{\epsilon }_{q^{\prime }\nu k_{z}}\cdot \mathbf{%
\chi }\right| ^{2},\text{ when }q^{\prime }\text{ is large}}{\frac{\pi
e^{2}q^{\prime 2}}{v}\left| \mathbf{\epsilon }_{q^{\prime }\nu k_{z}}\cdot 
\mathbf{d}\right| ^{2}\text{, when }q^{\prime }\text{ is small\ }},
\end{equation}%
where

\begin{equation}
\mathbf{d=}\int d^{2}\mathbf{\tau }w_{c}(\mathbf{\tau }_{z}\mathbf{,}\tau
_{\varphi })\tau w_{v}(\mathbf{\tau }_{z}\mathbf{,}\tau _{\varphi }).
\end{equation}

Eq. (30) can not be evaluated analytically. But for large radius, the
asymptotic form of Hankel function is : $H_{0}^{(1)}(q^{\prime }\rho )\sim 
\sqrt{2/(\pi q^{\prime }\rho )}e^{iq^{\prime }\rho }.$ And the renormalized
frequency shift reduces to 2D limit\cite{19}:

\begin{equation}
\Omega _{2D}=-\gamma _{\text{single}}(\frac{1}{k_{0}d})^{2},
\end{equation}
where

\begin{equation}
\gamma _{\text{single}}=\frac{2e^{2}}{\hbar c}\frac{E_{\nu ,k_{z}\sim 0}}{%
\hbar }\left| k_{0}\mathbf{d}\right| ^{2}
\end{equation}%
is the radiative decay rate of a single isolated exciton. Analogous to the
decay rate, the renormalized frequency shift is explicitly seen to be
coherently enhanced by the same factor $(1/k_{0}d)^{2}$ as a result of the
interaction of the phase-matched photon amplitude with the delocalized
excitonic amplitude in the plane. In Fig. 3, we numerically calculated the
frequency shift as a function of wire radius. One can see from the figure
the renormalized frequency shift does not show oscillatory dependence on
radius, either. For $\nu =0$ mode, unlike the behavior of decay rate, the
magnitude of the frequency shift first increases with the decreasing of
radius. After reaching a minimum point, the frequency shift approaches to
zero rapidly. 
\begin{figure}[th]
\includegraphics[width=8cm,clip=true]{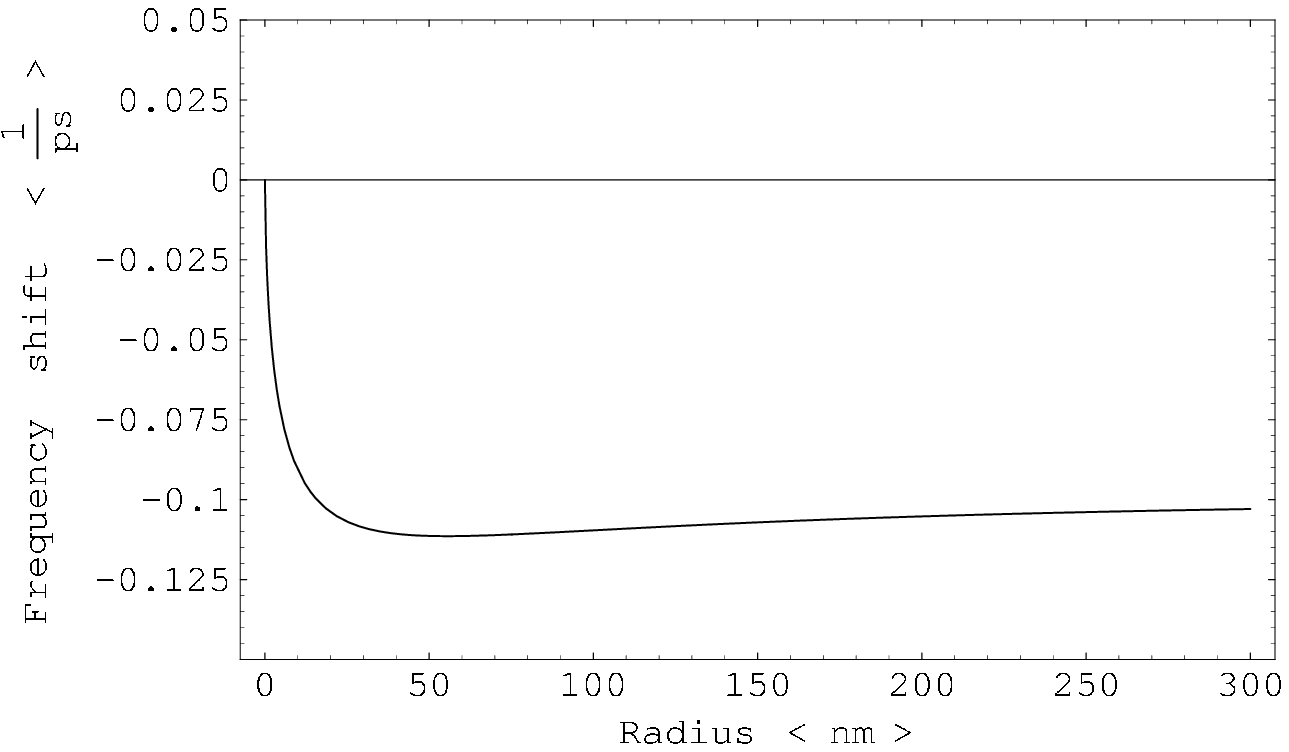}
\caption{{}Renormalized frequency shift of the superradiant exciton as a
function of the wire radius. The vertical and horizontal units are ps$^{-1}$
and $nm,$ respectively.}
\end{figure}

For usual semiconductors, the superradiant enhanced factor is about $10^{6}$
for excitons in the optical range. However, due to the extreme smallness of $%
\gamma _{\sin gle}$ itself, observation of $\Omega _{\nu ,k_{z}}$ is not
expected to be easy. Its dependence on wire radius may be a useful feature
to observe this quantity. Recently, R. A. R\"{o}mer and M. E. Raikh studied
theoretically the exciton absorption shredded by a magnetic flux $\Phi $ in
a quantum ring\cite{23}. They found the oscillator strength of the exciton
is most enhanced when $\Phi $ is equal to half of the universal flux quantum 
$\Phi _{0}=hc/e$. And the oscillation period is equal to $\Phi _{0}.$ Owing
to the similarity in cylindrical geometry, the decay rate and the frequency
shift may be observed experimentally if one varies the magnetic flux or the
wire radius.

In summary, we have calculated the decay rate of the superradiant exciton in
a hollow cylindrical quantum wire. Similar to the case in a quantum well,
the higher wave number modes are shown to have larger maximum decay rate. It
is also found the decay rate does not show any oscillatory dependence on
wire radius because of the conservation of the angular momentum in the
cylindrical system. On the other hand , the frequency shift of the exciton
is properly renormalized by removing the ultraviolet and infrared
divergence. It is found the shift does not shown oscillatory dependence on
the radius, either. Some distinguishing features are pointed out and may be
observed in a suitably designed experiment.

We would like to thank Professor M. F. Lin of National Cheng Kung University
for a helpful discussion. This work is supported partially by the National
Science Council, Taiwan under the grant number NSC 91-2112-M-009-012.

\end{document}